\documentclass[numbers]{elsarticle}
\usepackage{latexsym}
\usepackage{natbib}
\usepackage[latin2]{inputenc}
\usepackage[hidelinks]{hyperref}
\usepackage{longtable, lineno}

\makeatletter
\def\ps@pprintTitle{%
	\let\@oddhead\@empty
	\let\@evenhead\@empty
	\def\@oddfoot{\centerline{\thepage}}%
	\let\@evenfoot\@oddfoot}
\makeatother

\begin{document}

\title{Near Perfect Protein Multi-Label Classification with Deep Neural Networks}


\author[p]{Balázs Szalkai\corref{cor1}}
\ead{szalkai@pitgroup.org}
\author[p,u]{Vince Grolmusz\corref{cor1}}
\ead{grolmusz@pitgroup.org}
\cortext[cor1]{Joint corresponding authors}
\address[p]{PIT Bioinformatics Group, Eötvös University, H-1117 Budapest, Hungary}
\address[u]{Uratim Ltd., H-1118 Budapest, Hungary}

\date{}


\begin{abstract}
	Artificial neural networks (ANNs) have gained a well-deserved popularity among machine learning tools upon their recent successful applications in image- and sound processing and classification problems. ANNs have also been applied for predicting the family or function of a protein, knowing its residue sequence. Here we present two new ANNs with multi-label classification ability, showing impressive accuracy when classifying protein sequences into 698 UniProt families (AUC=99.99\%) and 983 Gene Ontology classes (AUC=99.45\%).

\end{abstract}

\maketitle

\section*{Introduction} 

Proteins are widely studied by numerous highly sophisticated tools in life science laboratories and by computational approaches. One important problem is the functional annotation or classification of proteins, using only structural information of these molecules. 

There are several levels of protein structure characterization \cite{Creighton1993}: primary, secondary, tertiary and quaternary structures. The primary structure describes the residue (i.e., amino acid) sequence; the secondary structure characterizes the regions of local, highly regular substructures, like $\alpha$-helices and $\beta$-sheets; the tertiary structure is the three-dimensional geometry of the folded substructures, while the quaternary structure describes the multi-subunit assembly of proteins, where each subunit consists of a single poly-peptide chain. 

Therefore, the most basic protein structure is the primary, while the most complex is the quaternary. It is well-known that hundreds of proteins with known quaternary  structures publicly deposited in the Protein Data Bank \cite{PDB-base} still lack satisfying functional annotation \cite{Nadzirin2012,pdbmend}.

With the knowledge of tertiary or quaternary protein structures of non-annotated proteins, sometimes it is possible to find small, characteristic parts of the molecules that may help their functional annotation. In enzymes, the chemical details of the active site \cite{Ivan2010} can be characteristic, as it was shown e.g. in the case of ASP-HIS-SER catalytic triads \cite{Ivan2009}, where the position of just four spatial points described the function of the enzyme well.

When only the primary structure, i.e., the residue sequence of the protein is known, it is more challenging to assign proper functions to these macromolecules. 

One possible approach is the sequence alignment-based similarity search between the input residue sequence $x$ and a properly chosen and functionally annotated reference sequence database $D$. For the sequence alignment one may use the exact Smith-Waterman algorithm \cite{Smith1981,Ivan2016}, or the popular BLAST or its clones \cite{Altschul1990}, or a more advanced, hidden Markov-model based HMMER search \cite{Eddy2009,Eddy2011,Szalkai2014}. Suppose that the similarity search for input $x$ returns the functionally annotated $y\in D$ as the most similar sequence from $D$. Then we may assign the function of $y$ to $x$. In other words, the input is assigned the function of the most similar sequence in a reference database. 

One deep problem with this simple sequence alignment approach is that the protein sequences have more and less conservative subsequences, and a similarity in the latter has less relevance than in the former. Another related problem is that the three-dimensional structure of the proteins are much more conserved in evolution than the primary structure. By some measures, defined in \cite{Illergaard2009}, they are three-to-ten times more conserved. Therefore, there could be big differences in the primary structures of two proteins with almost the same three-dimensional shape and with the very same function.

Consequently, more sophisticated classification methods are needed than the simple sequence alignment approach. 

\subsection*{Neural networks for protein classification}

A fast developing area of research is the application of artificial neural networks (ANNs) for protein classification. Artificial neural networks are perhaps the most widely used artificial intelligence tools today, frequently applied for classification and -- nearly real-time -- image- and sound processing for numerous applications, e.g., \cite{Srivastava2014,He2014,He2015,Ioffe2015}.  ANNs contain artificial neurons or perceptrons \cite{McCulloch1943} as basic building blocks,  each of which computes a non-linear function of the weighted sum of its inputs. This non-linear function is termed the activation function. Then the output of the neuron may be fed to another neurons as input. The neurons in the first layer (called the input layer) work on the input of the network. The output of the network is computed by the output layer. When the problem to solve is a classification task, each class is assigned a different neuron in the output layer, which is activated if the input is classified into the corresponding class. 

While the building blocks of neural nets are the artificial neurons (or perceptrons), they can be viewed as a set of neuron layers. The output of a layer becomes the input of the next layer or the output of the whole network (for the last layer). A layer is a parametric function with learnable parameters. The network is the composition of these functions, and itself can be thought of as a parametric function $f_\eta$ with $\eta$ as the weight-parameter vector that assigns the weights of the inputs of each artificial neuron in the network. Today's neural networks are mostly deep neural networks, meaning that they have a much larger number of layers than earlier variations, resulting in a vastly increased learning capacity.

If one specifies the non-linear activation functions of the neurons, and the architecture of the network, then the neural network can be trained to perform its classification task. This learning capability is the most appealing property of neural nets. The weights of the neuron inputs, i.e., the vector $\eta$ is improved step-by-step, as described below.

Neural networks are usually trained in a supervised fashion by inputting specific $x$ values into it and backpropagating the error $\hat{y} - f_\eta(x)$. This means that a loss function is applied on the network output $y = f_\eta(x)$ and the desired output $\hat{y}$, and the parameters of the network are updated with gradient descent. The network input $x$ can be modeled as a random variable, and $\hat{y} = g(x)$ is a function of $x$ that needs to be approximated by the network. Let $\varepsilon(y, \hat{y})$ denote a loss function (e.g., $\varepsilon(y, \hat{y}) = (y-\hat{y})^2$ is a possible choice, named the L2 loss function). Then the expected loss of the network can be written as

$$
\ell(\eta) = \mathbf{E}_x \varepsilon(f_\eta(x), g(x)),
$$

i.e., as a function of the network parameter vector $\eta$.

This formula can then be approximated by substituting the expected value for the mean over a given set of possible inputs mirroring the actual input distribution. If $x_1, ... x_n$ denote a random sample of inputs, then an approximation of the above formula is

\[
\tilde{\ell}(\eta) = \frac{1}{n} \sum_{i=1}^n \varepsilon(f_\eta(x_i), g(x_i)).
\]

Updating the network weights can be done with stochastic gradient descent (SGD) \cite{Amari1993} using the update step $\eta_{k+1} := \eta_{k} - \lambda \frac{\partial\tilde{\ell}}{\partial \eta} (\eta_k)$. The initial value $\eta_0$ is initialized randomly. If the learning rate $\lambda$ is sufficiently small, $\eta_k$ will then converge to a place of local minimum.

Stochastic gradient descent is susceptible to stalling near saddle points in the error surface, causing slow convergence. Furthermore, the size of the update steps (the differences in $\eta$) can be too large if $\lambda$ is too big, and this may result in divergence. Therefore SGD has since been improved by introducing momentum or using the statistics of the gradients to normalize the update steps and thus yielding the more modern methods such as RMSProp, Adagrad or Adam \cite{Kingma2014,Dauphin2015,Dauphinb}.

Classification problems with the inputs possibly corresponding to multiple classes are called multi-label classification problems. When we intend to classify proteins by using their amino-acid sequences as inputs and different functional or structural classes as outputs, we also need multi-label classification procedures, as a protein may be assigned one or more functional or structural classes.

\subsection*{Previous work}

The NNPDB program described in \cite{Wu1991} applied an n-gram model for classification into a small number of classes. In the work of \cite{Ferran1993} the protein sequences were stored in 20 x 20 bi-peptide matrices, and the neural network was trained in an unsupervised manner. The accuracy of the method was not reported explicitly. 

In \cite{Wu1993} the ProCANS tool was constructed for the classification of the members of the PIR database \cite{Barker2000}, by using the n-gram model with SVD (singular value decomposition) and MPL (multi-layer perceptrons). The proteins were encoded by the ae12 system into length-462 vectors. In the performance evaluation the authors of \cite{Wu1993} counted the classification as ``exact'' if the correct superfamily was present in the first 5 suggestions, i.e. in a quite ``tolerant'' way. Even with this definition, the family classification accuracy is 97.02\%, somewhat worse than our precision and recall values (where we say that a classification is successful when it returns the exact family of the protein). Additionally, in \cite{Wu1993}, for the computation of the singular value decomposition (SVD) the authors applied 659 training and also the 235 test proteins, and not only the 659 training proteins. The target classification classes in this work are pairwise disjoint, i.e., the authors did not solve a multi-label classification task (while we do in the present contribution).

The neural networks, built in \cite{Wu1995}, applied the n-gram model with SVD and MPL and attained a 90\% sensitivity when classifying unknown sequences into 3311 PIR superfamilies and families \cite{Barker2000}. In \cite{Wu1998} a hybrid neural network -- sequence alignment approach was applied for gene family identifications. Our approach does not use sequence alignment, just pure ANN tools in the classification.

The authors of \cite{Pasquier1999} classified proteins in transmembrane and not-transmembrane groups with ANNs. In a subsequent work, \cite{Pasquier2001} subdivided the non-transmembrane proteins into further three classes, i.e., four classes in total. Our contribution classifies the whole UniProt into 698 classes, and not just 4.

In the publication \cite{Wang2000a}, protein sequences were classified into four superfamilies only. 
The article \cite{Wang2002b} constructed a neural network for classifying 10 superfamilies from the PIR database \cite{Barker2000}.  The neural network constructed in \cite{Chen2003a} performed a yes-no classification into 2 classes: globin or non-globin.

Choosing input from the Protein Data Bank \cite{PDB-base}, \cite{Weinert2004} constructed and trained multiple fully-connected multilayer perceptrons (MLPs) for function prediction. The accuracy rate of the prediction was 75\%. The works of \cite{Blekas2003,Blekas2005} apply a hybrid motif-search \& neural network approach for classifying into a maximum of 7 protein classes.

Using a small network and two models: a simple 2-gram model (lsa2) and a more involved hydrophobicity-based (hyd2) model, the authors of \cite{Rossi2007} performed protein classification into five functional classes and four families. Further protein sequence classification results were reviewed in \cite{Saha2013} up to the year 2013. 

More recently, in \cite{Cao2014} proteins from the PIR database were classified into 10 superfamilies with a maximal accuracy of 93.69\%.  The work \cite{Chicco2014} described an ANN-based Gene Ontology functional classification solution that yielded less than 90\% AUC for one class, and 80\% AUC for two further classes; our Gene Ontology classification results have an AUC of 90.69\%.

In the article \cite{Cerri2015} Gene Ontology \cite{Consortium2015} classification was done with AUC values of around 0.5; our AUC values in the present contribution are around 0.9. We classify into 983 classes, while the authors of \cite{Cerri2015} into 2849 classes.

In \cite{Nguyen2016}, by applying convolution networks, DNA sequences were classified into a small number of classes (less than 10); our classifications use much more classes with high accuracy classification (in the case of UniProt, 698 classes). 

In the FFPred 3 tool, the authors of \cite{Cozzetto2016} trained separate SVM's for several hundred classes and attained F1 values under 43\%. Our F1 values are around 86\% in Gene Ontology and 98\% in UniProt classification. 

The work of \cite{Lee2016} performed UniProt classification with different methods (SVM, LSTM, GRU, CNN) and with 589 target classes. Their best F1 value is 94.8\%, while we classify into 698 classes with a better F1 value 98.63\% in the present work.

In the publication \cite{Liu2017}, the authors trained the neural network with 80\% of the sequences of the SwissProt subset of UniProt, and test the performance on the remaining 20\% of the sequences from SwissProt. The authors attain nearly 100\% accuracy, but they only classified into 4 classes, while here we show a neural network that classifies SwissProt into 698 classes with a near-100\% accuracy.

\section*{Methods and Materials}

We have applied the SwissProt subset of the UniProt protein database \cite{Consortium2009}, acquired from \url{http://uniprot.org} as starting point (using the query ``goa:(*) AND reviewed:yes''), containing 526,526 sequences having Gene Ontology IDs \cite{Consortium2015} at the date of download of 15 February 2017. The sequences were downloaded along with their assigned UniProt families. This set was shuffled and then divided into training and test sets using the bash commands \texttt{head -5000} and \texttt{tail -n +5001}. Since the data had headers, the test set contained 4,999 protein sequences, and the training set had the rest (521,527 sequences).

We trained two models on this dataset: one for Gene Ontology functional classification \cite{Consortium2015} and one for UniProt family classification \cite{Consortium2009}. Each of these had to solve a {\it multilabel classification task}, as a sequence could have been assigned more than one functions or families. There was a logical relationship among the attributes (functions/families) in both cases, describable using a directed acyclic graph (DAG), where each edge signifies an implication: for each edge $A \rightarrow B$, if an entry belongs to the class (has attribute) A, then it will also belong to the class B. Classes that do not have exiting edges are termed the {\em roots} of this graph, and the {\em level} of each class is a non-negative integer corresponding to the length of the shortest path from that class to a root. This means that roots have level 0, and each non-root node (class/attribute) has a level greater than 0. (c.f., Figure 1).

In Gene Ontology, the \texttt{is\_a} relation defines this directed acyclic graph on the functional attributes, e.g. if a protein has the function ``thyroid hormone generation'' (GO:0006590), then this implies that the protein falls into the category ``thyroid hormone metabolic process'' (GO:0042403), which in turn implies the functions ``phenol-containing compound metabolic process'' (GO:0018958), ``cellular modified amino acid metabolic process'' (GO:0006575) and ``hormone metabolic process'' (GO:0042445), because if a compound is a ``thyroid hormone'', then it is also a ``phenol-containing compound'', ``cellular modified amino acid'' and ``hormone''.

\begin{figure}
	\centering
	\includegraphics[width=5in]{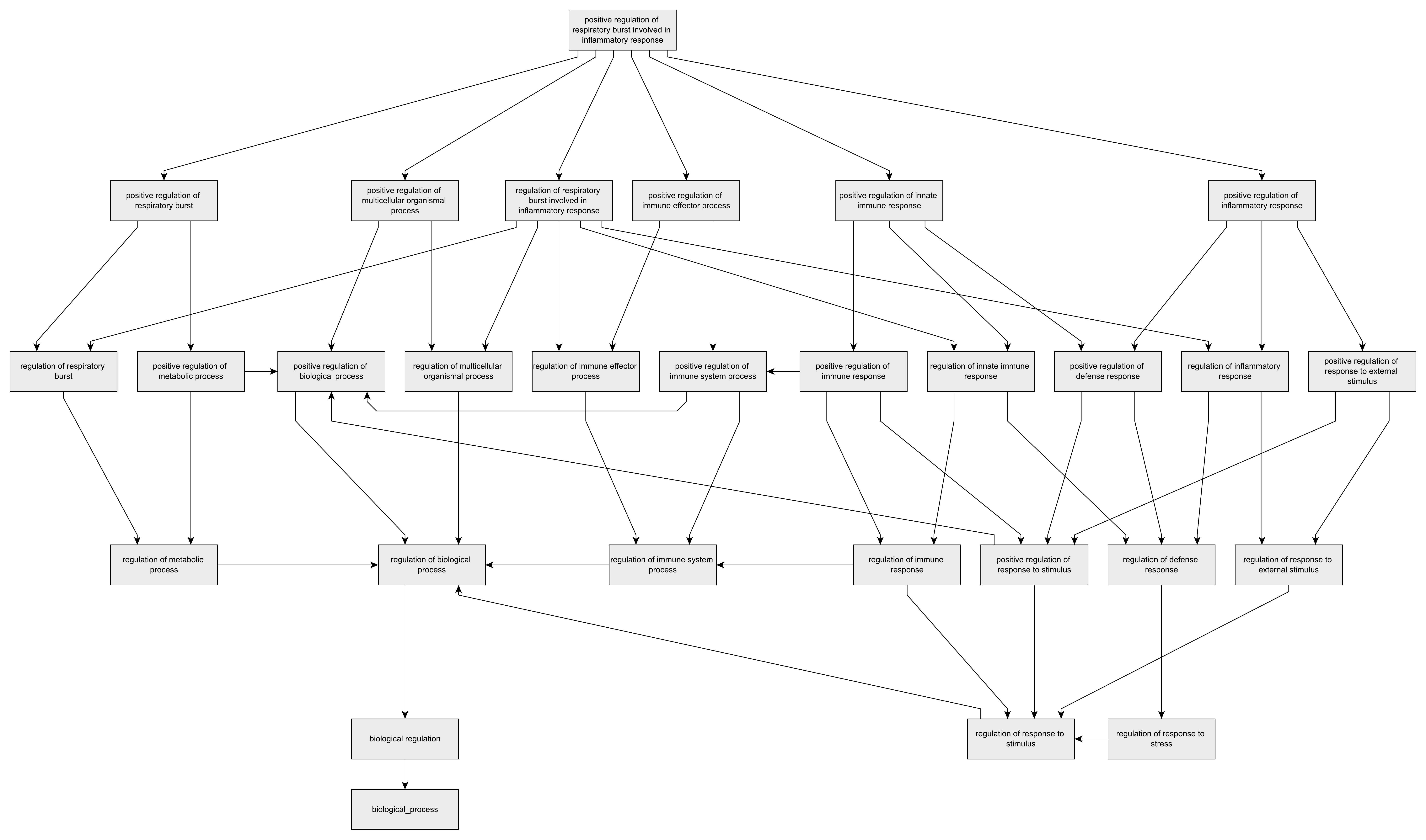}
	\caption{A small example from Gene Ontology \cite{Consortium2015}. Each edge corresponds to an implication: for each edge $A \rightarrow B$, if an entry belongs to the class (has attribute) A, then it will also belong to the class B. Classes that do not have exiting edges are termed the {\em roots} of this graph, and the {\em level} of each class is a non-negative integer corresponding to the length of the shortest path from that class to a root. }
\end{figure}

On the other hand, UniProt families exist on 4 levels: superfamily, family, subfamily and sub-subfamily. Here each class can belong to zero to one parent class, and may or may not have child classes, so the relationships of the classes can be represented by a forest of directed trees (arborescences). For example, ``HOG1 sub-subfamily'' belongs to ``MAP kinase subfamily'', which belongs to ``Ser/Thr protein kinase family'', which is a child of ``Protein kinase superfamily''. But not all roots of the directed forest are superfamilies: for example, ``Mimivirus L114/R131 family'' has no parent despite being a family and not a superfamily.

When training the network, each training protein sequence was fed as the input ($x$) of the network, along with the classification target ($\hat{y}$) of that sequence encoded as a 0-1 vector. Each class was represented as a coordinate in the target vector. If the sequence belonged to a specific class, then all the classes reachable from that class in the DAG were included in the classification target, encoded as ones in the target vector $\hat{y}$. If the sequence did not belong to a specific class (or any of its subclasses), then the corresponding component of the target vector was zero.

The input sequences were encoded as two arrays: one 3-dimensional array \texttt{inputSeq} with dimensions \texttt{[batch\_size, max\_length, dims]} and another array \texttt{inputSeqLen} encoding the length of the individual sequences with dimension \texttt{[batch\_size]}. Here \texttt{batch\_size} means the number of sequences in a minibatch and was set to 32. \texttt{max\_length} was the maximum allowed length of a sequence: sequences longer than this were omitted in the training phase and cropped to the first \texttt{max\_length} amino acids in the testing phase. This parameter was set to 2000 in our case. These parameters were largely determined by the available video memory on our GPU (4GB RAM). Parameter \texttt{dims} (=26) was the length of the vectors encoding the individual amino acids. We encoded each amino acid as a 26-dimensional vector, where the first 20 components comprised a one-hot vector (all components zero except the one uniquely identifying the amino acid in question), while the other 6 components encoded various properties of the amino acids: charge ($\pm 1$ or 0.1 in the case of Histidine which is positive about 10\% of the time and neutral 90\% of the time), hydrophobicity, and the binary attributes \texttt{isPolar}, \texttt{isAromatic}, \texttt{hasHydroxyl} and \texttt{hasSulfur}. Apart from this straightforward encoding scheme, we did not use any other information about the biological properties of the sequences or their amino acids, including the secondary structure or the presence of pre-selected motifs. This means that the neural network had to work with the amino acid sequence alone, without any further help from other machine learning methods.

At training time we confined the set of sequences to those between 162 (this was the minimum length for the neural network, so the output of the last pooling layer was at least one amino acid long) and 2000 (a practical limit because of available memory). The starting ``M'' (Methionine) character was removed from all the sequences. We also excluded those classes from the attribute graph which had fewer than 200 or 150 sequences in the training set in the case of Gene Ontology and UniProt family classification, respectively. The classes were considered up until level 3 in both graphs (in the case of UniProt families, this was not a real restriction, as the graph already had only 4 levels ranging from level 0 to level 3 in that case). All UniProt sub-subfamilies had too few members, so in fact, all the UniProt sub-subfamilies were dropped, leaving a total of 3 levels (0..2) in that case. In the end, 983 classes were considered in the Gene Ontology task and 698 classes in the UniProt family classification task.

The deep neural network had a primarily convolutional architecture with 1D spatial pyramid pooling \cite{He2014} and fully connected layers at the end. The architecture is shown in Table~\ref{table:network}. The network had 6 one-dimensional convolution layers with kernel sizes \texttt{[6,6,5,5,5,5]} and depths (filter counts) \texttt{[128,128,256,256,512,512]}, with PReLU (parametric rectified linear unit) activation \cite{He2015}. We used max pooling with kernel size and stride 2 after each convolutional layer, except the first one. Max pooling was omitted after the first layer so that the network can conserve details about the fine structure of the protein. Each max pooling layer was followed by a batch normalization layer to help normalize the statistics of the heatmaps.

After the last convolutional layer, we applied a 1D variant of SPP (spatial pyramid pooling) to convert the output of the last max pooling layer into a fixed-length representation of each (variable-length) sequence. We performed SPP on 3 levels with 1, 4 and 16 divisions, respectively. This means that the activation of the neurons was max-pooled over the whole sequence, then the sequence was divided into four almost equally sized parts and the activations were max-pooled over each of the 4 subsequences, then again the sequence was divided into 16 parts, yielding 21 values altogether for each sequence and each of the 512 filters. Consequently, after SPP, the network state could be represented as an array of shape \texttt{[batch\_size, 21, 512]}.

\begin{table}
	\label{table:network}
	\begin{tabular}{|c|}
		\hline
		\textbf{Gene Ontology functional classifier network} \\ 
		\hline
		conv (size=6, stride=1, depth=128, padding=VALID, activation=prelu) \\ 
		\hline
		batch\_norm (scale=False) \\ 
		\hline
		conv (size=6, stride=1, depth=128, padding=VALID, activation=prelu) \\ 
		\hline
		max\_pool (size=2, stride=2, padding=VALID) \\ 
		\hline
		batch\_norm (scale=False) \\ 
		\hline
		conv (size=5, stride=1, depth=256, padding=VALID, activation=prelu) \\ 
		\hline
		max\_pool (size=2, stride=2, padding=VALID) \\ 
		\hline
		batch\_norm (scale=False) \\ 
		\hline
		conv (size=5, stride=1, depth=256, padding=VALID, activation=prelu) \\ 
		\hline
		max\_pool (size=2, stride=2, padding=VALID) \\ 
		\hline
		batch\_norm (scale=False) \\ 
		\hline
		conv (size=5, stride=1, depth=512, padding=VALID, activation=prelu) \\ 
		\hline
		max\_pool (size=2, stride=2, padding=VALID) \\ 
		\hline
		batch\_norm (scale=False) \\ 
		\hline
		conv (size=5, stride=1, depth=512, padding=VALID, activation=prelu) \\ 
		\hline
		max\_pool (size=2, stride=2, padding=VALID) \\ 
		\hline
		batch\_norm (scale=False) \\ 
		\hline
		spp (levels=3, divs\_per\_level=4) \\ 
		\hline
		fully\_connected (units=1024, activation=prelu) \\ 
		\hline
		dropout (p=0.5) \\ 
		\hline
		batch\_norm (scale=True) \\ 
		\hline
		fully\_connected (units=983, activation=sigmoid) \\ 
		\hline
	\end{tabular}
\end{table}

The output of the spatial pyramid pooling layer was fed into a fully-connected layer with 1024 units and PReLU activation, followed by a dropout layer with $p=0.5$ to avoid overfitting \cite{Srivastava2014}, and a batch normalization layer to normalize the mean and standard deviation. Then a second fully connected layer with sigmoid activation assigned numerical values (likelihoods) between 0 and 1 for each class, yielding the output array $y$ with shape \texttt{[batch\_size, n\_classes]}. Note that softmax activation cannot be used because the network had to perform a multi-label classification task.

We defined the loss of the neural network as the weighted cross entropy between the predictions and the targets. If $C$ denotes the number of classes, and $\ell$ the network loss, and $N$ the minibatch size, then let

\[
\ell := \frac{1}{N} \sum_{i=1}^N \sum_{j=1}^C w_j \hat{y}_{ij}(-\log y_{ij}) + (1 - \hat{y}_{ij})(-\log(1 - y_{ij})).
\]

Here the class weights $w_j$ ($1\leq j\leq C$) are responsible for class balancing to avoid misclassification of instances belonging to infrequent classes. $w_j$ is defined as $\max\{1, \min\{5, \mu(s) / s_j\}\}$, where $s_j$ is the size (number of sequences) of the $j$th class, and $\mu(s)$ is the mean of the class sizes. Thus the error of misclassifying an instance originally belonging to an infrequent class is weighted up by a factor between $1$ and $5$.

We also added an L2 regularization penalty to $\ell$ to reduce the risk of overfitting, with $\lambda = 6 \times 10^{-11}$.

Both neural networks (the Gene Ontology and also the UniProt family classifier) were implemented in TensorFlow \cite{Abadi2016a,Abadi2016c,Rampasek2016},  and trained for 150000 iterations (minibatches), i.e. 9.2 epochs, with a fixed learning rate of $0.002$. Training took 29 hours on an NVIDIA Geforce GTX 750 Ti GPU. For simplicity, no validation set was used, as overfitting was hoped to be largely addressed by the regularization methods (batch normalization, dropout, L2 regularization). We calculated various performance measures of the networks on the test set, including precision, recall and F1-value, both per class and altogether. The AUC (area under the ROC curve) was calculated using micro-averaging (the ROC curve is the true positive rate as the function of the false positive rate). 

For each level of the class graph, the perfect prediction rate was also determined: this was defined as the number of sequences where the set of classes on the specific level of the graph was perfectly predicted by the network. As the test set had sequences shorter than the minimum length or longer than the maximum length, the networks could not be tested on all the test sequences, but only 3776 and 3744 sequences, respectively.

\section*{Results and Discussion}

The performance of the two networks on the test set is summarized in Table~\ref{table:performance} below. To our knowledge, both of our networks outperform all previously described purely neural solutions on these classification tasks, detailed in section ``Previous work'' in the Introduction. 

In the evaluation of the results we apply the following quality measures: 

{\em Precision} denotes the number of true positives divided by the number of predicted positives.

{\em Recall} (or sensitivity) denotes the number of true positives divided by the actual number of positives. 

The {\em F1 score} is the harmonic mean of precision and recall. 

The area under the ROC curve ({\em AUC}) corresponds to the probability that a model outputting a score between 0 and 1 ranks a randomly chosen positive sample higher than a randomly chosen negative sample.

For example, our UniProt family classifier network achieved an F1-value of $98.63\%$, much better than the $94.85\%$ reported by \cite{Lee2016} (which was achieved in an easier task, involving a classification into only 589 instead of 698 classes).

\begin{table}
	\label{table:performance}
	\begin{tabular}{lccccc}
		\hline
		Network & \# classes & Precision & Recall & F1-value & AUC \\
		\hline
		Gene Ontology & 983 & 91.17\% & 81.62\% & 86.13\% & 99.45\% \\ 
		\hline
		UniProt families & 698 & 99.75\% & 97.53\% & 98.63\% & 99.99\% \\
		\hline
	\end{tabular}
	\caption{The general evaluation of the performance of our neural networks. Note the very high number of classes and the near-perfect accuracy, compared to the previous work, listed in the Introduction. {Precision} denotes the number of true positives divided by the number of predicted positives. {\em Recall} (or sensitivity) denotes the number of true positives divided by the actual number of positives. The {\em F1 value} is the harmonic mean of precision and recall. The area under the ROC curve ({\em AUC}) corresponds to the probability that a model outputting a score between 0 and 1 ranks a randomly chosen positive sample higher than a randomly chosen negative sample.}
\end{table}

We also calculated the per-level precision and recall of the two networks. Probably because of the smaller number of nodes and greater difference among protein sequences in different classes, the topmost level is the easiest to classify for the Gene Ontology network. The UniProt family network performed the best on level 2 (UniProt subfamilies that have a containing family and superfamily), but this is probably because there were only 13 nodes at that level.

\begin{table}
		\centering
	\begin{tabular}{|c|c|r|r|r|}
		\hline
		Level & Classes & Precision & Recall & F1-value \\ 
		\hline
		0 & 3 & 93.99\% & 98.23\% & 96.07\% \\ 
		\hline
		1 & 53 & 89.60\% & 85.00\% & 87.24\% \\ 
		\hline
		2 & 236 & 91.01\% & 80.78\% & 85.59\% \\ 
		\hline
		3 & 691 & 91.42\% & 77.16\% & 83.68\% \\ 
		\hline
	\end{tabular}
	
	\caption{Per-level performance of the Gene Ontology network}
\end{table}

\begin{table}
		\centering
	\begin{tabular}{|c|c|r|r|r|}
		\hline
		Level & Classes & Precision & Recall & F1-value \\ 
		\hline
		0 & 524 & 99.84\% & 97.68\% & 98.75\% \\ 
		\hline
		1 & 161 & 99.31\% & 96.66\% & 97.97\% \\ 
		\hline
		2 & 13 & 100.00\% & 100.00\% & 100.00\% \\ 
		\hline
	\end{tabular}
	
	\caption{Per-level performance of the UniProt family network}
\end{table}

From our results and previous work we can conclude that the Gene Ontology functional classification task seems to be harder for artificial neural networks than the UniProt family classification task, probably because the assignment of UniProt families depends heavily on sequence similarity, and thus it is easier to classify proteins into UniProt families instead of functional classes based purely on the amino acid sequence data. Additionally, the class graph has an easier (forest) structure in the UniProt family case.

The ROC curves for the two classifier networks are shown in Figures 2 and 3. From the AUC values (99.45\%, 99.99\%) it is clear that the networks achieve excellent classification performance, unmatched by prior architectures, listed in the Introduction.

\begin{figure}
	\includegraphics[width=5in]{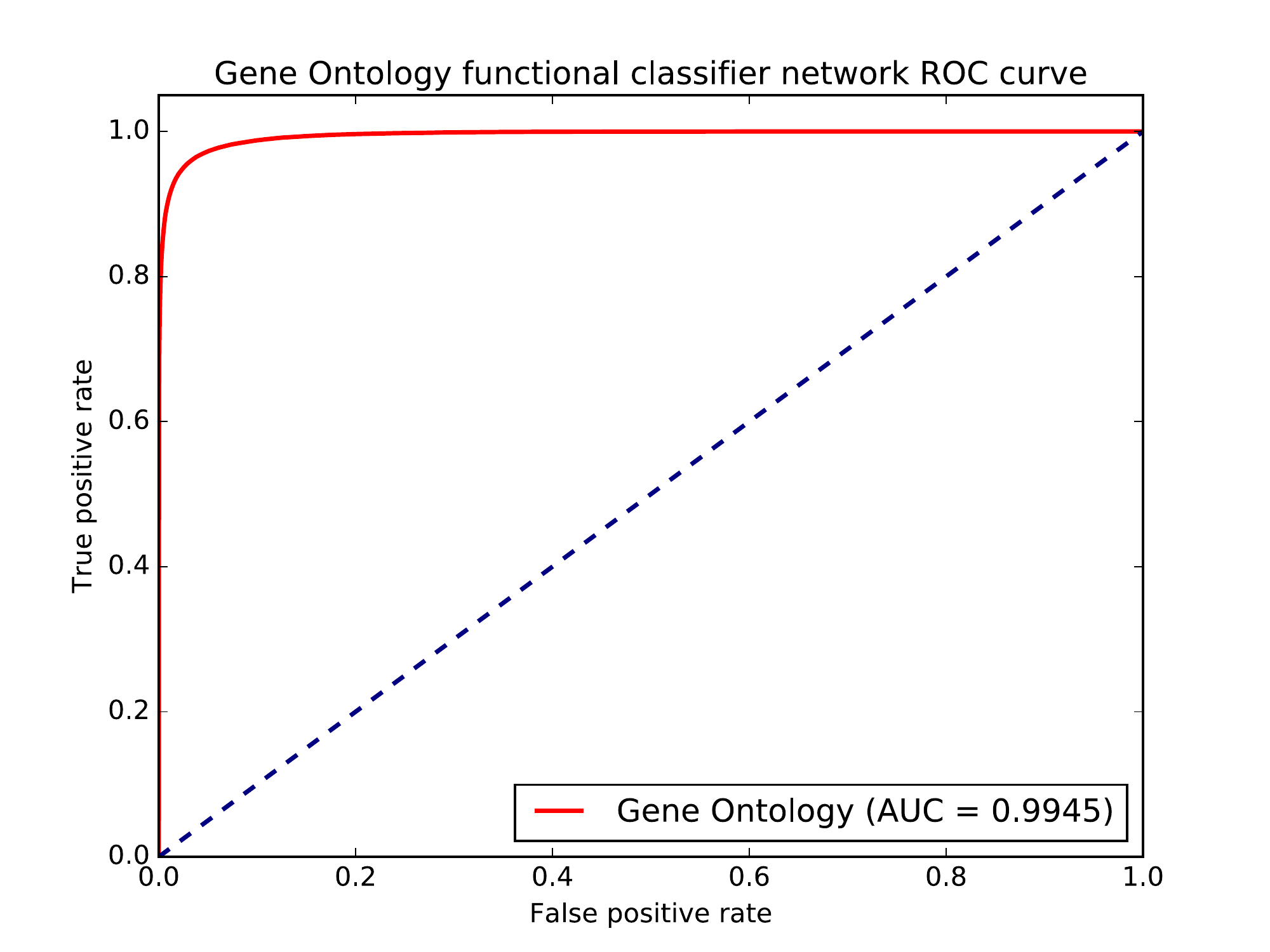}
	\caption{The ROC curve for the classifier network for the Gene Ontology task.The AUC value is 99.45\%}
\end{figure}

\begin{figure}
	\includegraphics[width=5in]{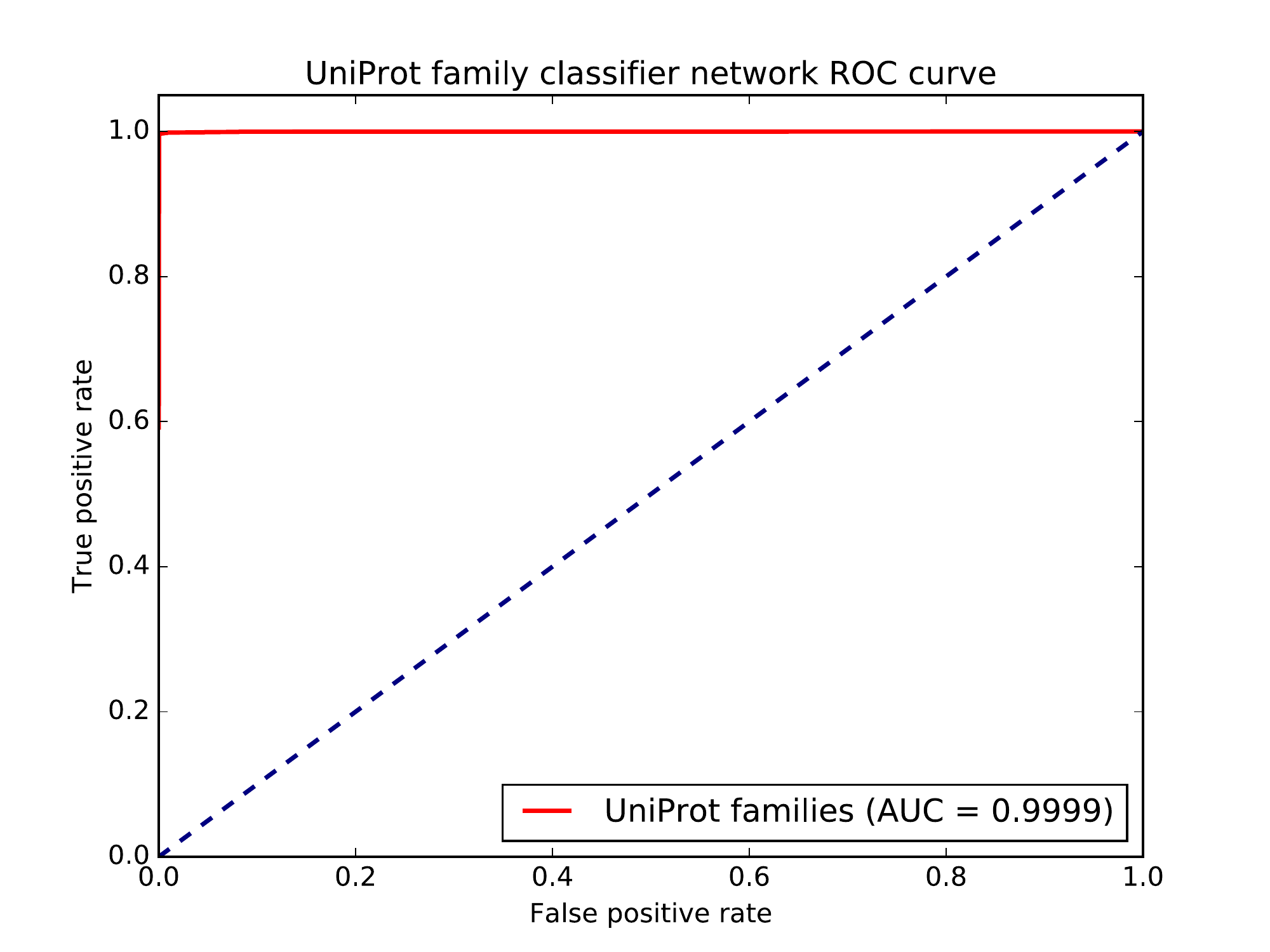}
	\caption{The ROC curve for the classifier network for the UniProt task. The AUC value is 99.99\%}
\end{figure}

\section*{Conclusions}

We have constructed deep artificial neural networks for protein classification into UniProt families and Gene Ontology classes. By the detailed comparison of previous work, our neural networks outperformed the existing solutions and have attained a near 100\% of accuracy in multi-label, multi-family classification.

 We also have conducted some experiments with the simplification of the network architecture. According to our experience, batch normalization is crucial to the performance of these networks, along with the number of layers (the overall depth of the networks): network variants without batch normalization and 5 (instead of 6) layers showed a performance drop of several percentage points. This emphasizes that deeper neural networks with more parameters have a much larger capacity for learning good representations, and normalizing the statistics of the layers can greatly improve network performance, as others already observed in image classification tasks \cite{Ioffe2015}. We hypothesize that, with more GPU RAM available, one can further improve upon the performance of our neural network by simply increasing the number of convolutional or fully connected layers, but overfitting may become a problem for such large network architectures.

\section*{Acknowledgments}
 BS was supported through the new national excellence program of the Ministry of Human Capacities of Hungary. The authors declare no conflicts of interest.



\end{document}